# Floware: Balanced Flow Monitoring in Software Defined Networks

Luiza Nacshon
Information Systems Engineering
Ben-Gurion University of the Negev
Beer-Sheva, Israel
luiza@post.bgu.ac.il

Rami Puzis
Information Systems Engineering
and Telekom Innovation Laboratories
Ben-Gurion University of the Negev
Beer-Sheva, Israel
puzis@bgu.ac.il

Polina Zilberman
Computer Science
and Telekom Innovation Laboratories
Ben-Gurion University of the Negev
Beer-Sheva, Israel
polinaz@cs.bgu.ac.il

## ABSTRACT

OpenFlow is a protocol implementing Software Defined Networking, a new networking paradigm, which segregates packet forwarding and accounting (performed on switches) from the routing decisions and advanced protocols (executed on a central controller). This segregation increases agility and flexibility of a networking infrastructure and reduces its operational expenses. OpenFlow controllers expose standard interfaces to facilitate variety of networking applications. In particular, a monitoring application can use these interfaces to push into the OpenFlow switches rules that collect traffic flow statistics at different aggregation levels. The aggregation level determines the monitoring accuracy and the induced network overhead. In this paper, we propose Floware – an OpenFlow application that allows discovery and monitoring of active flows at any required aggregation level. Floware balances the monitoring overhead among many switches in order to reduce its negative effect on network performance. In addition, Floware integrates with monitoring systems based on legacy protocols such as NetFlow. We demonstrate the application with soft switches emulated in Mininet, the Floodlight controller, and the NetFlow Analyzer as a legacy network analysis and intrusion detection system. Evaluation results demonstrate the positive impact of balanced monitoring.

## 1. INTRODUCTION
Software Defined Networking (SDN) simplifies network management and makes it more flexible, responsive to the changing requirement and sudden protocol updates. In conventional networks both the data-plane and the control-plane are managed by the same network device. In SDNs, however, the control-plane is implemented by a remote software-based controller. Due to this segregation, SDN devices are simpler, cheaper, and more efficient than regular network devices and require less firmware updates. The agility, flexibility, and lower operational expenses of SDN make it a natural solution for the highly dynamic cloud networks [1].

This paper focuses on collecting traffic flow statistics in OpenFlow – a common protocol that implements SDN [2][3]. OpenFlow provides a few basic mechanisms for flow monitoring: polling the switches for statistics stored in its flow-table entries (pull-based) and letting the switch report the statistics to the controller (push-based approach), see Section 2 for details. Both mechanisms consume network resources and their careless and pervasive usage can reduce the network performance [11].

OpenFlow switches store traffic flow statistics in flow-table entries. Such entry can match one flow between specific source-destination addresses (exact-match) or an aggregated flow. For example, flow-table entries installed for routing purpose usually represent aggregated flows in order to save network resources. Aggregations reduce the monitoring accuracy along the IP address-space dimension (see Section 3 for details). In contrast, it is not feasible to install an exact-match flow-table entry for every pair of IP addresses in the monitored IP range. Excessive flow-table entry installation has many negative effects, such as redundant control messages, errors, and packet drops.

The primary objective of this study is providing flow monitoring with the highest granularity in the IP space while reducing its negative effects on the network. We introduce Floware, an OpenFlow application that manages and optimizes the flow monitoring process. Floware provides the following unique features:

**1. Active flow discovery:** Floware efficiently discovers new active flows that match a specific condition. Throughout this paper, we consider flows that pass through a given set of switches, but same techniques can be used to discover active flows matching other conditions as well.

**2. Balanced flow monitoring effort:** Given the set of aggregated flows, Floware assigns each flow to a switch where the flow can be monitored with minimal negative effects on the network. Floware collects the statistics using the push-based approach, but same techniques can be utilized to optimize polling of statistics as well.

**3. OpenFlow network accessibility for legacy traffic analysis tools:** Floware enables upgrading the network fabric to OpenFlow without the need to replace legacy flow monitoring infrastructure, which traditionally supports NetFlow, sFlow, IPFIX and similar protocols. The interoperation of OpenFlow with legacy traffic analysis systems is demonstrated using NetFlow Analyzer.

The rest of the paper is structured as follows: In Section 2 we provide the essential background on monitoring in OpenFlow and discuss the associated tradeoffs in Section 3. Section 4 presents the main algorithmic and technical contribution including the general architecture, the process of discovering active flows, and the balanced flows assignment. A showcase of Floware that includes integration with NetFlow Analyzer as well as the evaluation of the balanced flow assignment is presented in Section 5. In Section 6 we discuss related works including qualitative comparison to Floware and integration opportunities. In Section 6, we bring the paper to a close with conclusions and implications for future research.

## 2. OPENFLOW ESSENTIALS

In this section, we proceed with a short review of OpenFlow mechanisms that are utilized in this paper. A reader familiar with OpenFlow control messages may choose to skip this section.

### 2.1. Flow-table entries

Similar to NetFlow, an OpenFlow flow-table entry is uniquely identified by a set of key fields. In addition to the NetFlow key fields OpenFlow flow-table entry also includes the following key fields: source MAC; destination MAC; Ethernet type; and VLAN ID. All fields can be masked to allow aggregating flows. Similar to the NetFlow cache, and to routing tables in regular networks, partial aggregation is allowed on the source IP and destination IP fields.

In addition to the key fields, every flow-table entry (also known as a flow handling rule) contains statistics, priorities and actions. Similar to NetFlow, the OpenFlow flow-table entry maintains packets and bytes counters. Every incoming packet is looked up in the OpenFlow flow-table. In the case of a match, packet and byte counters of the matching flow-table entry are updated. Since this operation is an inherent part of the OpenFlow design and often performed in the hardware, no additional resources (CPU or RAM) are required (as opposed to NetFlow's design). Next, the switch acts according to the matching flow-table entry action field. This field can direct the switch to do one of the following: drop the packet, look it up in the next flow-table in the pipeline (OpenFlow v1.3), send it to a controller, or output it to a port.

If a packet matches several entries, the entry with the highest priority field value is selected. In the case that a packet matches two entries with the same priority, the entry with the closest match is used. In case that no matching flow-table entry is found for the incoming packet, the switch encapsulates the packet in a packet-in message and sends it to the controller [2].

### 2.2. The OpenFlow control plane

The main function of the control plane is to define the packet routes by setting the action field of the flow-table entries. There are two methods for installing flow-table entries: reactive and proactive. With the former approach, the controller reacts to packet-in messages by determining the routing path for the new flow. If the routing path has not yet been calculated, the controller may flood the network with a packet-out message in order to discover an available route. The controller sends flow-mod messages to all the switches along the determined route in order to install flow-table entries for forwarding consequent packets. A flow-mod message contains the data necessary to populate all the flow entry fields, including, first and foremost, the action field.

Most OFCs support custom modules that enable the network administrator to alter the controller's response to control messages. These modules are organized in a pipeline where every incoming message is processed by each module in a row. These modules can define the order in which they process the control messages. They also are able to allow or disallow further message propagation along the pipeline – a feature that we extensively utilize in Floware.

In addition to the reactive installation of the flow-table entries described above, the controller may install the entries proactively. This method allows the network administrator to install flow-table entries before receiving the packet-in message.

The network administrator may set the timeouts of the installed flow-table entries. An entry may have two types of timeouts: hard timeout – when the flow entry will expire after the defined time period; and idle timeout – when the flow entry will expire if no matching packet is received during the defined period of time. These two timeouts are similar to the active and inactive timeouts in NetFlow. Flow-table entries with no timeouts never expire and are considered as static flow-table entries.

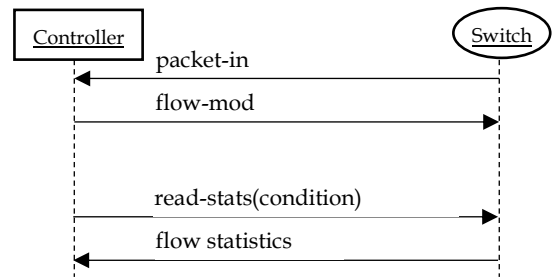

**Figure 1. OpenFlow pull-based monitoring**

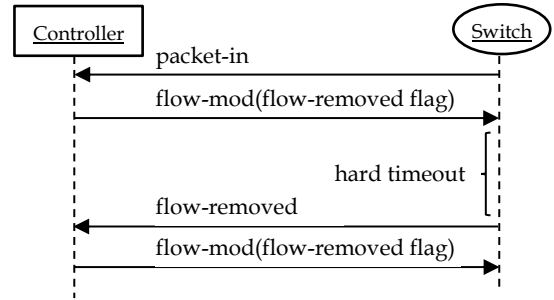

**Figure 2. OpenFlow push-based monitoring**

In this paper we use both proactive and reactive installation of flow-table entries as explained in Section 4.

### 2.3. Flow monitoring

There are two approaches for collecting flow statistics in OpenFlow: pull-based and push-based (also referred to as active and passive monitoring respectively).

In the pull-based approach (see Figure 1), the controller sends a read-stats message to the switch requesting statistics of flows matching a given condition. The condition can match a single flow-table entry or a number of them. Pull-based monitoring is not suitable for continuous high-granularity monitoring. It is said to consume too great a portion of switch-controller bandwidth and switch CPU, limiting the use of these resources in flow setup for routing purposes [11]. For example, Sunnen [20] showed that when the read-stats messages are sent too often, the switch's CPU utilization and the number of the pending messages increases.

In contrast, push-based monitoring (see Figure 2) is passive and requires less network resources [21]. According to this approach, switches send the controller a flow-removed message with statistics about the expired flow. The switch can be configured to report statistics on selected flows by setting the flow-removed flag on each flow that needs to be monitored. The frequency of reporting is determined, in this case, using the hard and idle timeout fields.

For each flow, the network administrator can choose a single switch that will contain the respective flow-table entry with the flow-removed flag set. Since flow-table entries matching this flow on other switches do not report statistics, redundant reporting is avoided. In this paper we employ push-based monitoring because of its lower switch-controller message overheads.

## 3. THE MONITORING TRADEOFFS

Accurate bandwidth provisioning, anomaly detection, and network health monitoring require collecting flow statistics at multiple points in the network. The collected flow statistics are analyzed, for example, to detect anomalous traffic patterns that can indicate a malfunction or a cyber attack.

In OpenFlow networks, statistics can be collected, either by polling the switches, or by installing flow-table entries with the flow-removed flag set, as described in Section 2. Next, we discuss several tradeoffs between network resource consumption and monitoring accuracy associated with the push-based monitoring approach. In general, there are two kinds of resources that need to be considered when planning a monitoring strategy: free flow-table entries and control messages.

**The time dimension:** Collecting flow statistics involves flow-removed control messages that are sent when a flow expires. A network administrator may utilize existing flow-table entries (installed for routing purposes) to also report flow statistics. In this case, flow-removed messages would be the only overhead [21].

One of the drawbacks of such an approach is the low granularity of the collected statistics both in the time and the IP-space dimensions. The statistics which are collected only upon flow expiration (a single sample per flow) may be insufficient for detecting attacks. Some monitoring frameworks, such as PayLess [22], sacrifice additional control messages in order to increase the time granularity of the statistics collection.

**The address-space dimension:** Some OFCs, such as OpenDayLight, install aggregated flows for routing purposes. For instance, consider flows with wild-carded source addresses. One one hand, wild-carded flow-table entries reduce the number of control messages and the load on the flow-table. On the other hand, it reduces the monitoring accuracy along the address-space dimension due to the aggregated source addresses. For example: NIDS attack detection accuracy may drop if the suspicious attack patterns are aggregated together with legitimate flows; oversized flows may be blended among many lightweight flows damaging route optimizations; etc. Measuring statistics of flows between specific sources and destinations requires installing flow-table entries with no aggregations (referred to as exact-match flow-table entries). In this case, the monitoring overhead consists of packet-in, flow-mod, and flow-removed messages as well as excess flow-table entries.

It is not feasible to install exact-match entries for all possible traffic flows in a network because they will soon fill up the flow-tables. When the switch's flow-table is filled it will respond with full-flow-table error for every flow-mod message [2]. This will consume the switch-controller bandwidth; increase the controller memory and CPU usage; cause routing vagaries as the controller tries to find alternative paths; and increase jitter and packet loss [11],[20],[23]. It is, therefore, extremely important to reduce the number of installed flow-entries. In this paper we propose a Flow Discovery technique to install exact-match entries only for active flows that need to be monitored.

Overall, there is a clear tradeoff between the resources spent on monitoring and the granularity of the collected statistics on both the time and the address-space dimensions. More statistics require additional flow-mod messages to install flow-table entries and additional flow-removed messages to collect the statistics. Superfluous control messages consume switch-controller bandwidth, memory/CPU of the switch and the controller, leading to delays and packet drops [20].

## 4. FLOWARE FRAMEWORK

In this section we present the Floware framework that enables the integration of legacy flow-based monitoring systems with Software Defined Networks (SDN). Floware includes a set of components for discovering active flows in the network, balancing the network resources used for collecting statistics, and exporting the collected statistics to an external monitoring system.

In regular networks this activity is facilitated by a variety of flow monitoring protocols such as NetFlow, JFlow, sFlow, IPFIX etc. Multiple NetFlow-Enabled routers export statistics on traffic flows passing through them to NetFlow-collectors that analyze and visualize the collected data to the network administrator [13], [14]. In order to maintain the routine of network monitoring, we allow network administrators to define a set of *observed switches* (OBS) and designate a NetFlow collector. The NetFlow collector will receive statistics on all flows passing through these devices at the highest granularity level as if the statistics were collected using NetFlow enabled on these devices. Unfortunately, the individual flows whose statistics need to be collected are not known a priori. Therefore, we propose a new Flow Discovery technique that requires only few additional control messages and flow-table entries distributed wisely across the network to avoid overload. We refer to these flow-table entries as flow-discovery entries throughout the paper.

### 4.1 Floware Synopsis

In this subsection we present the general idea of our approach toward flow monitoring and the conceptual architecture of Floware. The monitoring approach presented in this paper is summarized in Algorithm 1. Accordingly, Floware first selects the routes passing through the OBSs (line 1). Floware generates aggregated static flow-discovery entries for routes selected in line 1 in order to discover new active flows (line 2). Floware installs the flow-discovery entries such that the monitoring load is equally balanced across the network switches (line 3). The action field of the flow-discovery entries is set to send to controller. Once the entries are installed Floware listens to packet-in messages triggered by new active flows (line 4).

Figuratively speaking, flow-discovery entries are used to trap active flows. An active flow is trapped when its first packet matches the flow-discovery entry. When this happens, the switch generates the packet-in message. Then, Floware receives it, and reacts by installing exact-match flow-table entries for the newly discovered active flow in order to collect statistics (lines 5-6). The timeouts of active flows and hence the frequency of the statistics collection is determined in lines 5 and 10 by a pluggable scheduling algorithm. We recommend employing an adaptive scheduling algorithm provided by PayLess [22]. For the purpose of inclusiveness we describe its simplified version in Section 3.4. Active flows are installed with a flow-removed flag set (lines 6 and 11). The action field of an active flow entry instructs the switch to forward the packet according to the routing strategy used in the network.

When Floware receives a flow-removed message generated due to the expiration of an active flow it first extracts the flow statistics (line 8), generates a NetFlow datagram and sends it to the NetFlow Collector (line 9).

Figure 3 depicts the conceptual architecture of Floware. Its main modules are responsible for: (a) generating the relevant flow-discovery entries (b) assigning them to switches (c) scheduling the expiration of active flows and (d) exporting flow statistics to the remote flow analyzer.

The Flows Discovery module generates the aggregated flow-discovery entries by selecting the routes passing through the OBSs and determining the source and target address spaces at each endpoint (e.g. ingress/egress routers). The endpoints, their subnets, and the routes between the endpoints are retrieved from the controller (interaction 2 in Figure 3).

The Flows Assignment module is responsible for balancing monitoring load across the network switches. Based on the capacities and occupation of switch flow-tables, it instructs the Flows Discovery module as to where each flow-discovery entry should be installed (interaction 3 in Figure 3). The Flow Assignment algorithm is detailed in Section 3.3.

The Scheduler is responsible for installing entries for active flows and scheduling their expiration (i.e., the monitoring frequency) in order to collect high granularity statistics (interactions 6 and 8 in Figure 3).

Data Export module listens to flow-removed messages from the active flows installed by the Scheduler (interaction 7 in Figure 3), generates corresponding NetFlow datagrams and sends them to a remote NetFlow Collector (interaction 9 in Figure 3).

In the rest of this section we elaborate on the functionality of each module, the interactions between Floware and controller, and the monitoring process from the switch perspective. The three major parts of the monitoring process are depicted in Figure 4.

## 4.2 Flows Discovery

During the first stages of the monitoring commencement (see Figure 4.a) Floware analyzes the underlying network in order to select routes passing through the OBSs and to generate the respective flow-discovery entries as explained in this subsection.

Information about switches and links can be extracted via the Northbound API of a Controller. For example, Floodlight [24] provides this functionality through a REST API[1]. Similarly it is possible to query the OpenFlow Controller for endpoints (ingress/egress switches)[2] and routes between them[3]. In this paper we consider an endpoint such as an access switch as being both ingress and egress.

Let G(V,E) denote the network topology where V is the set of switches and E is the set of links between them. Let S⊆V and T⊆V be the sets of source and destinations switches respectively. Every traffic flow enters the network through a source switch s∈S and leaves the network through a destination switch t∈T.

We denote by $IP(v) = \{IP_1, \dots, IP_n\}$ the set of IP subnets that communicate with the network through the endpoint v∈S∪T. Given an source switch s∈S and a destination switch t∈T, we distinguish

**Algorithm 1. Floware Monitoring Process**

*Flow Discovery*:
1. Select the routes passing through the OBSs
2. Generate aggregated *flow-discovery* entries
3. Install the aggregated *flow-discovery* entries
4. Listen to packet-in messages

**On packet-in** (due to flow-discovery entry on switch R):
5. Set the monitoring frequency of the *active flow*
6. Install the exact match entry for the *active flow* on switch R
7. Listen to *flow-removed* messages

**On flow-removed** (due to expiration of active flow f):
8. Extract flow statistics from f
9. Export NetFlow datagram
10. Update monitoring frequency of the *active flow f*
11. Reinstall the *active flow f* on the same switch.

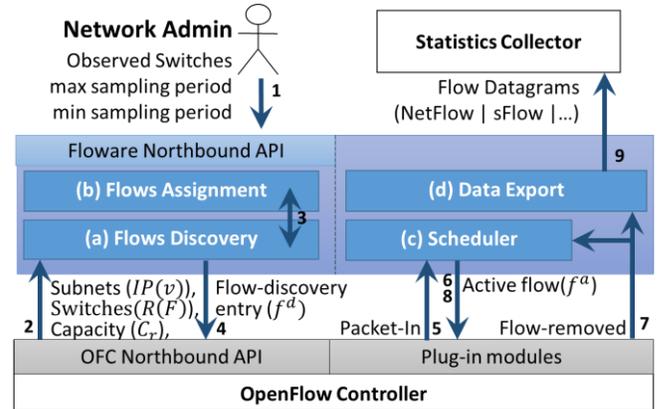

Figure 3. Floware architecture

between two types of flows: aggregated $(IP_i, IP_j)$, where $IP_i \in IP(s) \wedge IP_j \in IP(t)$, and exact-match $(ip_k, ip_l)$ where $ip_k \in IP_i$ and $ip_l \in IP_j$. For the sake of simplicity, in the rest of this paper we ignore other flow attributes such as protocol type, ToS, etc. We define F as a set of aggregated flows between all pairs of source/destination switches:

$$F = \{(IP_i, IP_j) \mid IP_i \in IP(s) \wedge IP_j \in IP(t) \wedge s \in S \wedge t \in T\}$$

Let $R: F \to 2^V$ denote the function which maps a flow $f \in F$ to its route $\{s, v_1, \dots, t\} \subseteq V$ within the network. Although in general, routes are ordered sequences of switches, we disregard the order in this paper. We generate flow-discovery entries for a subset of aggregated flows $F^d \subseteq F$ whose routes pass through at least one of the OBSs:

$$F^d = \{f^d \in F \mid R(f^d) \cap OBSs \neq \emptyset \} \quad (1)$$

**Example 1:** Consider the sample network presented in Figure 5. There are three switches: one source, one destination, and one functioning as both source and destination. The destination switch is selected as the so called OBS in order to monitor flows passing through it. Each switch is associated with one or two IP spaces and the set F represents all possible flows within the network. The flow from x.y.w.b/29 to x.y.z.a/26 is routed through the direct link between the two right switches. Thus, only four out of the five flows in F pass through the OBS (denoted by $F^d$).

---

[1] /wm/core/controller/switches/json

[2] /wm/device/

[3] /wm/topology/route/src_router/src_port/dst_router/dst_port/json

Alternative flow selection and aggregation conditions are possible as well. For example, a network service provider (NSP) may choose to inspect flows targeted at clients subscribed to its managed security service. In addition, an intrusion detection system may focus on analyzing communication flows sharing similar destinations, payload, and platform on the source machine. Yen and Reiter [25] show that aggregations performed on such a focused excerpt of flows enable accurate botnet detection. Given the sets of source and destination switches (S and T respectively) and the OBSs defined by the network administrator, the Flow Discovery module generates and installs static flow-discovery entries as summarized in Algorithm 2. Line 1 initializes the set of flow-discovery entries as well as the map of flows to OBSs through which the flows pass. The FlowsToOBS map may later be required by the Data Export module as explained in Section 3.5. Next, in lines 2-3, we iterate over all subnets connected to all source and destination switches. A flow-discovery entry is generated for each pair of subnets in line 4 and saved for future use only if at least one of the OBSs is along its route (lines 5-6). We also save the OBSs where each flow could have been monitored for later use in line 7.

In line 8, the Flows Discovery module invokes the Flows Assignment algorithm to determine the location of each flow discovery entry. The result of Flow Assignment is a function $D: F^d \to V$ that maps flow-discovery entries to switches. Each generated flow-discovery entry is installed on the assigned switch (see lines 9-10 in Algorithm II, Figure 4.a, and interaction 4 in Figure 3). Finally, the two maps, that (1) define for each flow on which OBS it could have been collected (FlowToOBS) and (2) where it should be collected in the OpenFlow network (D), are transferred to Data Export module.

Each flow-discovery entry $f^d = (IP_i, IP_j)$ represents an aggregation of flows between machines within the subnets $IP_i$ and $IP_j$. Usually only few of these flows are simultaneously active. In order to discover these flows Floware sets the action field of the installed flow-discovery entries to send to controller and listens to incoming packet-in messages through the controller's native API.

A new active flow that matches a flow-discovery entry, denoted as $f^a \in f^d$, triggers a packet-in message on the switch where $f^d$ is installed. This message is received by the Scheduler (see Figure 4.b) through the native API of the controller (see interaction 5 in Figure 3). The Scheduler reactively installs exact-match active flow entries in response to the packet-in messages as will be detailed in Section 3.4.

At this point it is important to note that $f^a$ must be installed on the same switch as the flow-discovery entry that triggered the respective packet-in message. This is done in order to prevent packets, from the same flow triggering additional packet-in messages.

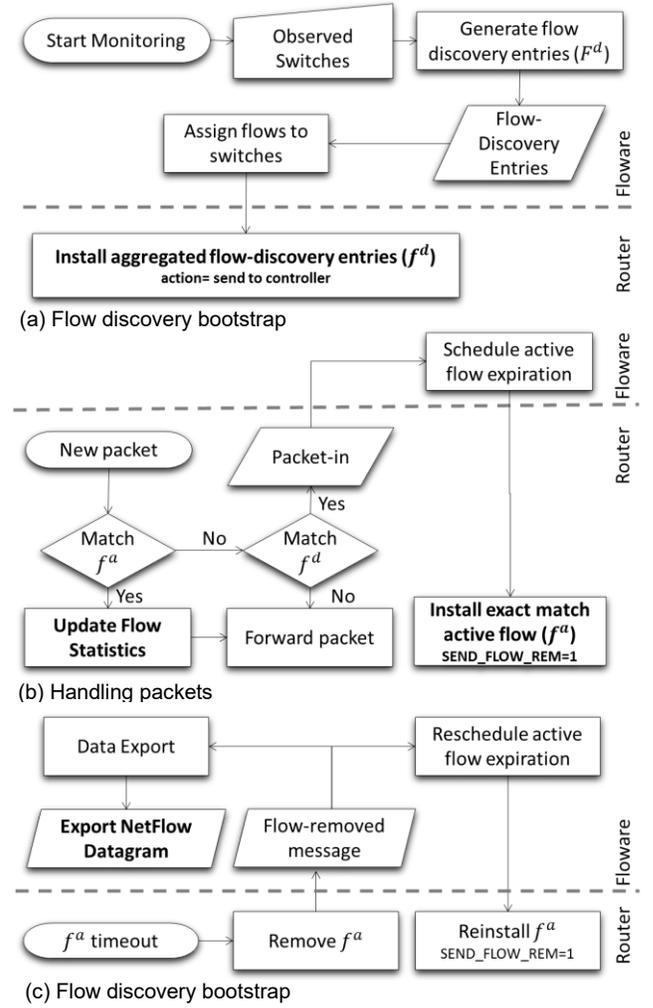

Figure 4. The monitoring process from the Floware and the switch perspectives

**Algorithm 2. Flows Discovery**

**Input:** $S, T, OBSs$

1. $F^d \leftarrow \emptyset, FlowToOBS \leftarrow \emptyset$
2. **For each** $s \in S$, $t \in T$
3.   **For each** $IP_i \in IP(s)$, $IP_j \in IP(t)$:
4.     $f^d \leftarrow (IP_i, IP_j)$
5.     **If** $OBSs \cap R(f^d) \neq \emptyset$:
6.       $F^d \leftarrow F^d \cup \{f^d\}$
7.       $FlowToOBS(f^d) \leftarrow OBSs \cap R(f^d)$
8. $D \leftarrow FlowsAssignment(F^d, R)$
9. **For each** $f^d \in F^d$:
10.   Install $f^d$ on $D(f^d)$
11. Send $FlowToOBS$ and $D$ to $DataExport$

We also note that Flow Discovery introduces an additional delay during initiation of monitored flows. When the first packet matching a flow-discovery entry arrives and triggers a packet-in message, the traffic flow is not immediately forwarded to the destination. The traffic forwarding continues after the active flow entry is installed.

Installing exact-match active flow entries significantly increases the number of flow-table entries installed on a switch. As explained in Subsection 2.2.4, an overfull flow-table causes error messages when controller attempts to install new flow-table entries and creates congestion at the overloaded switch. Therefore, it is very important to balance the monitoring load across the network switches in order to minimize the chance of exceeding the flow-table capacity as discussed in the next subsection.

### 4.3 Flow Assignment

In this subsection we discuss the load that monitoring creates on switches and propose a greedy algorithm to balance this load among the network switches instead of overloading few OBSs.

The Flow Assignment module is responsible for choosing the switches on which flow-discovery entries, generated by the Flow Discovery module, should be installed. Every flow-discovery entry ($f^d$) results in the installation of a number of exact-match active flow entries ($f^a \in f^d$) on the same switch. We denote the number of active flow entries that match the flow-discovery entry $f^d = (IP_i, IP_j)$ as $load(f^d)$. Let $\mu$ denote the expected fraction of active flows out of all possible flows matching $f^d$. The expected load created by $f^d$ is

$$load(f^d) = 1 + \mu \cdot |IP_i| \cdot |IP_j| \qquad (2)$$

Where $|IP_i|$ and $|IP_j|$ are the number of addresses in the $IP_i$ and the $IP_j$ subnets respectively. The unity in Equation 2 represents the flow-discovery entry and $\mu \cdot |IP_i| \cdot |IP_j|$ is the expected number of active flows that match $f^d$.

Note that, although $\mu$ may vary considerably for various aggregated flows, for the sake of simplicity, we refer to the fraction of active flows between any two subnets as $\mu$ without additional indices or parameters. If required, $\mu$ can be efficiently estimated for all pairs of source/destination switches using periodical snapshots of switch flow-tables or Traffic Matrix estimation techniques [26]. Prior research indicates that $\mu$ value is quite small in enterprise networks. For example, Naous et al. [27] collected results from the Stanford Computer Science and Electrical Engineering network [28] and found that with 5,500 active hosts the number of active flows at any second never exceeded 10,000—even at busy times.

Efficient distribution of flow-discovery entries balances the load on switches across the network such that no switch is overloaded. In this paper we employ a simple yet efficient greedy algorithm to balance load on switches (see Algorithm 3). The algorithm receives as the input the set of flow-discovery entries ($F^d$), computed in lines 1-6 of Algorithm 2, and the routes of the respective flows ($R: F^d \to 2^V$). Balancing the monitoring load relies on the number of free flow-table entries ($C_r$) in each candidate switch ($r \in V$) (lines 1-2). The number of free and used flow-table entries can be extracted from the controller Northbound API.[4]

Next, the algorithm iterates over all flow-discovery entries in the order of non-increasing load (lines 3-4). Each entry ($f^d$) is assigned to the switch along its path ($R(f^d)$) that has the maximal number of free flow-table entries (lines 5-6). The number of free flow-table entries is updated based on the expected load (see Equation 2) on the chosen switch in line 7.

---
[4] /wm/core/switch/all/table/json

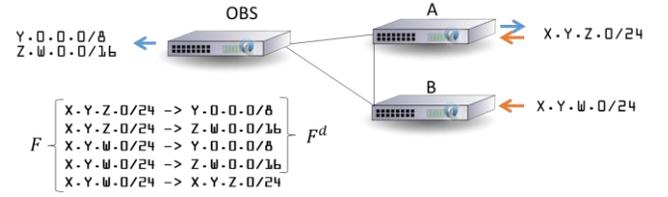

Figure 5. Aggregated flow-discovery entries

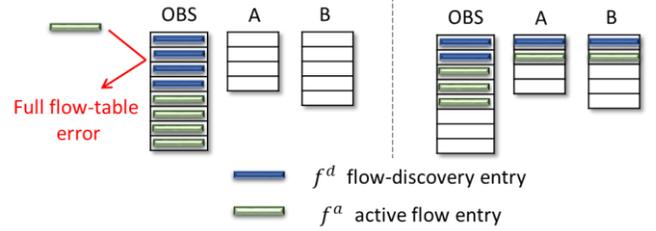

Figure 6. Unbalanced flow assignment (left) vs. balanced flow assignment (right).

**Algorithm 3. Flow Assignment**

| |
|---|
| **Input:** $F^d$, $R: F^d \to 2^V$ |
| **Output:** $D: F^d \to V$ |
| 1.     **For each** $r \in V$: |
| 2.         $C_r \leftarrow$ number of free flow-table entries in switch $r$ |
| 3.     Sort $F^d = \{f^d\}$ in the order of non-increasing $load(f^d)$ |
| 4.     **For each** $f^d \in F^d$ do: |
| 5.         $r \leftarrow ARGMAX_{x \in R(f^d)}\{C_x\}$ |
| 6.         $D(f^d) \leftarrow r$ |
| 7.         $C_r \leftarrow C_r - load(f^d)$ |
| 8.     **Return** $D$ |

Note that correct functioning of Flow Assignment as described here relies on the estimation of the expected fraction of active flows ($\mu$) and the estimation of the number of free flow-table entries for each candidate switch. Note also, that we assume in Algorithm 3 that there are enough free flow-table entries to install at least the flow-discovery entries. The algorithm will still function correctly if the number of free flow-table entries is smaller than the expected number of active flow entries that may be installed there. In such cases errors will be reported by the switches during later stages. But using the Flow Assignment algorithm that balances the load reduces the number of such errors as we show in Section IV.

**Example 2:** Once again, consider the sample network presented in Figure 5. There are four aggregated flows passing through the OBS. Their respective flow-discovery entries are represented by the dark bars in Figure 6. Intuitively, one may install all the flow-discovery entries on the OBS itself. Assume, for example, that there are five discovered active flows. As stated in Subsection 2.2, active flow entries must be installed on the same switch. Although, OBS has the most free flow-table entries (eight in Figure 6), it is not sufficient to accommodate both the flow-discovery entries and the active flow entries. The excess active flow entries will result in full flow-table errors as depicted in Figure 6 (left). Assume now that we assign the flow-discovery entries to several different switches: one to A, one to B, and two to the OBS. Since some active flow entries will now be installed on the switches A and B, the OBS will no longer be overloaded as depicted in Figure 6 (right).

## 4.4 Scheduler

Following the installation of flow-discovery entries, as described in Section 3.2, the Scheduler listens to packet-in messages triggered by the flow-discovery entries and installs respective exact-match active flow entries with the flow-removed flag set (see Figure 4.b). The Scheduler also listens to flow-removed messages triggered by the expiration of the installed active flows and re-installs these flows with adapted timeouts (see Figure 4.c).

The main objective of the Scheduler is to adapt the expiration frequency of active flows to ensure: 1) the collection of high granularity statistics and 2) minimal bandwidth consumption (reflected by the number of flow-mod and flow-removed messages). We propose using PayLess [22] to adapt the timeouts of active flows to dynamicity of the collected statistics. If the statistics (packets and bytes counters) collected for some active flow are characterized by high variability over time, this flow is re-installed with a decreased timeout. In the opposite case, the active flow is re-installed with an increased timeout. The minimal and maximal timeouts are determined by the network administrator (interaction 1 in Figure 3).

Upon the receipt of a packet-in message, triggered by a flow-discovery entry ($f^d$), the Scheduler installs an exact-match active flow entry ($f^a$) for the flow indicated in the packet-in message. $f^a$ is installed on the same switch where $f^d$ has been installed, but with higher priority than $f^d$. The action field of $f^a$ instructs the switch to forward matching packets according to the routing strategy used in the network. Packets matching $f^a$ update the flow-table entry's counters and are forwarded to the defined output port[5].

When the active flow entry expires the entry is removed, its statistics are encapsulated in a flow-removed message according to OpenFlow specification [2]. The message is sent to the controller. The controller passes the message to the Scheduler through the native API (see interaction 7 in Figure 3 and Figure 4.c).

## 4.5 Data Export

Data Export is the last module in the monitoring process. It is responsible for transferring the collected statistics to the remote NetFlow Collector. As explained in Section II, both the NetFlow cache and the OpenFlow flow-tables contain statistics on flows. In addition, both NetFlow and OpenFlow support push-based monitoring. Hence, the Data Export module can push the data collected by exact-match active flow entries to the remote collector (see interaction 9 in Figure 3 and Figure 4.c). The Data Export module extracts statistics data from flow-removed messages triggered by active flows expiration and converts the data to NetFlow v5 datagrams. See Table 1 for the detailed conversion map.

Note that NetFlow collectors (such as flow-based NIDS) run on a remote server and receive NetFlow records traditionally exported using User Datagram Protocol (UDP). The Data Export sets the destination address of the UDP packets to the IP address of the NetFlow collectors. Originally, the source address of the NetFlow datagrams should be the IP address of the OBS interface from which the statistics were collected. For the sake of flow analyzers that utilize this information, Floware can set the source address of the exported datagrams such that either: (1) the changes in the monitoring process are fully transparent to the NetFlow Collector; or (2) the collector receives accurate information with respect to the location were the statistics were actually collected.

In the first case, the Data Export module groups the flows according to the OBSs through which they could pass, and exports each group with the source address set to the respective OBS. To set this IP address correctly the Data Export module maintains a map between the flows in $F^d$ and the OBSs through which they pass. This FlowToOBS map is computed by the Flow Discovery module (see line 7 in Algorithm II).

In the second case, the exported datagrams contain statistics of flows that were installed on the same switch. The Data Export module sets the source address of the datagrams to the IP of the switch where the respective flows were installed.

## 5. FLOWARE EVALUATION
### 5.1 Evaluation Environment and Showcase

In this subsection we present the main components of the Floware showcase and the evaluation environment: the OpenFlow controller, network, and a traffic analysis tool.

**OpenFlow Controller.** Floodlight [24] is an open source, Java-based controller that is supported by developers from the Big Switch Networks company. Floodlight interacts with the network switches through its Southbound API using the OpenFlow v1.0 protocol. Figures 7 and 8 show the Floodlight Web UI. Floodlight provides the Northbound REST API to enable interaction with external applications. Through this API, it provides information about the routing, statistics, devices, features, etc. We implemented the Flow Discovery module of the Floware prototype as a Python application that interacts with Floodlight through the REST API. The Scheduler and the Data Export were implemented as custom modules that interact with Floodlight through the Java API.

**Network.** We experimented with networks emulated using Mininet V2.1.0 [29], a rapid prototyping platform for creating software-defined networks on a single machine. It operates the OpenVswitch software switches and uses localhost virtual Ethernet links between the switches. Mininet also provides built-in topologies and a Python-based API to build custom topologies. We used the latter feature to build topologies obtained from the RocketFuel project [30]. Figure 9 displays the Mininet console where the host h1 is instructed to ping the host h4.

**Traffic analysis tool.** In order to demonstrate the integration of NetFlow-based traffic analysis with SDN, we used ManageEngine's NetFlow analyzer with an advanced security analytics module (ASAM) add-on, version 9. ASAM generates traffic reports based on the NetFlow datagrams exported from OBSs. According to the information in the NetFlow datagrams, ASAM analyzes the traffic passing through OBSs and provides an overall security snapshot of the network [12]. The Data Export module generates NetFlow v5 datagrams that are sent to the preconfigured IP address and UDP port 9996 of the remote NetFlow Analyzer server. Figure 10 shows an example of an exported datagram that contains infromation on the traffic flow between the host h1 and the host h4 (having the IP addresses 10.0.0.1 and 10.0.0.4 respectively). Figure 11 depicts a report generated by NetFlow Analyzer based on flow statistics received from NFO.

We demonstrated Floware and the evaluation environment on an Ubuntu OS v 13.10 64Bit server with core×4 processor and 16GB memory. Once the Floodlight controller starts running, we build the network topology in Mininet with the Mininet Python-API. Then, once Mininet is connected to the controller, the controller learns the network topology and the links between the switches.

---

[5] The value of the output port is retrieved from the OFC through the Floodlight's native API: IRoutingDecision.getRoutingAction()

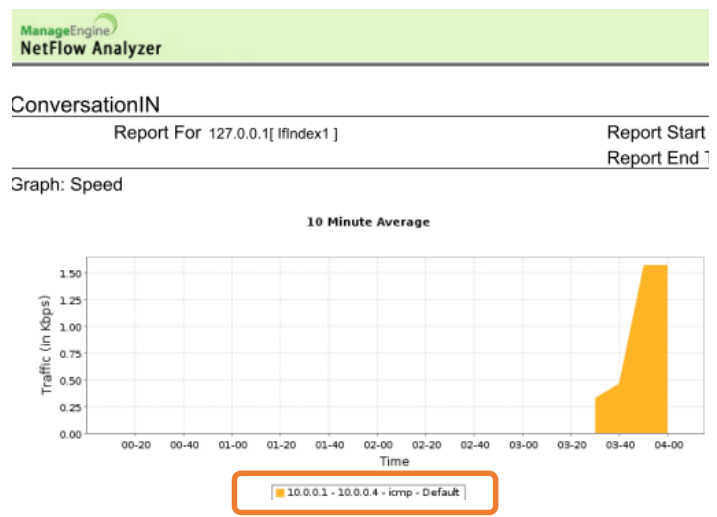

Figure 7. Active flow entries installed on a router.

Figure 8. IP and MAC adresses of routers in Floodlight.

Figure 9. The ping command in Mininet.

Figure 10. Wireshark capture of the NetFlow datagram exported by Floware.

Figure 11. NetFlow Analyzer receiving the data exported by Floware.

## 4.2 Experiment Setup

In this section we describe the experimental evaluation of Floware. We focus on evaluating the effect of flow assignment strategies on Floware performance. Two flow assignment strategies are considered: the greedy flow balancing algorithm as described in Section 3.3 (denoted as Balanced) and the baseline strategy where flow-discovery entries are installed on the OBSs (denoted as Baseline). Note that, in the Baseline strategy, when a flow-discovery entry can be mapped to multiple OBSs we randomly choose one of the OBSs on which to install the entry. This is done in order to allow fair comparison of the strategies with respect to the number of installed flow-discovery entries.

Chowdhury et al. [22] extensively evaluated the performance of PayLess as a scheduling algorithm and demonstrated its apparent benefits. However, the PayLess scheduler introduces additional noise into the evaluation of flow assignment strategies due to the changing frequency of active flow installation. In order to factor out the effect of the

Scheduler on the load created by monitoring, we use a baseline scheduler that sets the timeout of every installed active flow entry to 60 seconds as recommended by the NetFlow analyzer support team. Flow-discovery entries never expire and the timeouts of flows installed by the controller in order to route traffic are kept at their default value.

The evaluation was performed with 11-switches' and 37-switches' tree topologies generated by Mininet. In order to show that Floware performs well also on more complex topologies we include the AS-1755 (EBONE, Amsterdam) and the AS-4755 (VSNL India) topologies obtained from RocketFuel [30]. The former contains 15 switches and the latter 31 switches. In our simulations, each one of the switches was connected to ten virtual machines. These ten virtual machines were assigned IP addresses within a unique /28 subnet.

Every simulation was executed for 300 seconds. The simulation execution was split into cycles of 1 to 10 seconds. In order to simulate communication between virtual machines, during each cycle every virtual machine continuously pinged ten random peers. In order to fairly compare between evaluation scenarios, we used the same random seed for choosing the set of ping destinations. Since the timeouts of flow-table entries are constant, the shorter the flows, the more load they create on the switches. When flows are short-leaved (e.g. cycle=1 sec) new flow entries are installed before the old ones expire.

The larger the flow-tables, the more entries they can accommodate before generating full flow-table errors. We experimented with flow-tables of 300 to 3000 entries. Although, there are products using larger tables, in current experimental settings 3,000 entries are enough to handle all flows. Floware performance was evaluated with 1, 2, and 3 randomly selected OBSs. Once OBSs were chosen, the Flow Discovery module generated flow-discovery entries for the flows which were intended to pass through at least one OBS. Flow discovery entries were assigned to switches and installed after the network was built and the virtual machines started pinging each other in order to let the controller learn the network.

## 4.3 Measured parameters

During the experiment, we recorded the number of flow-table entries that were installed (denoted as total flow entries) including flow-discovery entries, active flow entries, and other entries installed by the controller. Intuitively, the network entries were not uniformly distributed across the network switches. Some switches were more heavily loaded than others due to their central position or traffic vagaries. The load on the switches can become even more dispersed if the monitoring load is not well-balanced. We used the Gini coefficient [31] to measure the dispersion of free flow-table entries in the network switches.

Occasionally, flow-tables become overfull especially when they are small. To capture the impact of overfull flow tables we measured the number of full flow-table errors. In order to obtain deeper insights into network performance during monitoring, we measured the number of packet-in messages separately for monitoring and for routing purposes (denoted as routing packet-in messages and monitoring packet-in messages respectively). Routing packet-in messages also included packet-in messages sent for ARP and any other network health check.

Every installed flow-table entry, except the static flow-discovery entries, should eventually be removed. Routing flow-table entries installed by the controller are removed without generating the flow-removed messages. However, the active flow entries installed by Floware do generate these messages. We measured the number of flow-removed messages as a proxy to the amount of collected statistics.

Excess control messages also consume the controller resources as pointed out by Tootoonchian et al. [32]. In this experiment we measured the memory usage of Floodlight controller.

## 4.4 Results

The Floware performance evaluation results are presented in Figures 12-19. We analyzed Floware performance from different perspectives and compared two flow assignment strategies: Baseline and Balanced. A qualitative comparison of Floware to related works is presented in Section V.

Figure 12 suggests that balancing the monitoring load across switches using the greedy flow assignment algorithm greatly reduces the chance for full flow-table errors compared to using only the OBSs for monitoring. Although this result is intuitive, it stands in contrast to the common practice of network monitoring where the fewest possible switches are selected to cover as many flows as possible [18],[19]. Full flow-tables also increase the number of control messages used for monitoring as well as for packet routing. Packet-in messages are used to notify the controller that a flow-table entry needs to be installed in order to handle this packet and all further packets from the same flow. However, if the flow-table entry is not installed, since the flow-table is full, further packets trigger additional packet-in messages consuming switch-controller bandwidth, CPU, memory, etc. For example, Figure 13 depicts the correlation between packet-in messages and full flow-table errors.

To better understand the relation between effective flow assignment and the effect of flow balancing on the network, in Figure 15 we plotted the total number of packet-in messages as a function of the Gini coefficient. We can see that the more balanced the distribution of free flow-table entries is (smaller Gini coefficient) the less redundant packet-in messages are in the network.

Figures 14 and 16 present the simulation results as the function of flow-table size and flow duration respectively. With a balanced distribution of flow records, it was possible to completely avoid errors (and excess control messages) with only 900 entries in the flow-tables of the switches in our experiment (see Figure 14). However, when the statistics are collected only from the OBSs, these switches need at least 2,400 entries in their flow-tables. In addition to saving switch resources, the proposed monitoring optimization saves controller resources as can be seen from the lower memory consumption of Floodlight (see Figure 17).

Furthermore, the greedy Flow Assignment algorithm proposed in this paper enables the installment of more flow-table entries for monitoring purposes as depicted in Figure 18. Thus more flow statistics are collected (see Figure 19) which increases monitoring accuracy along the IP space dimension.

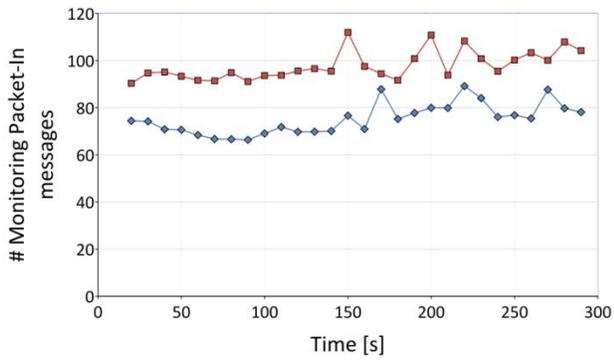

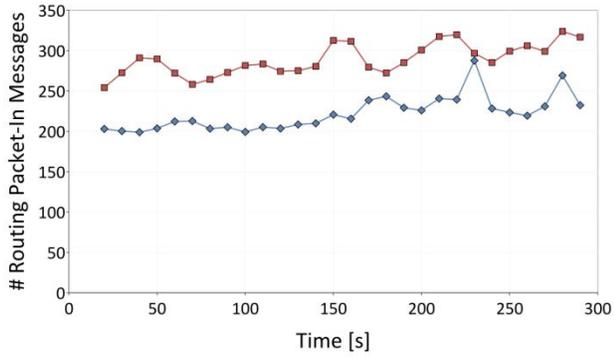

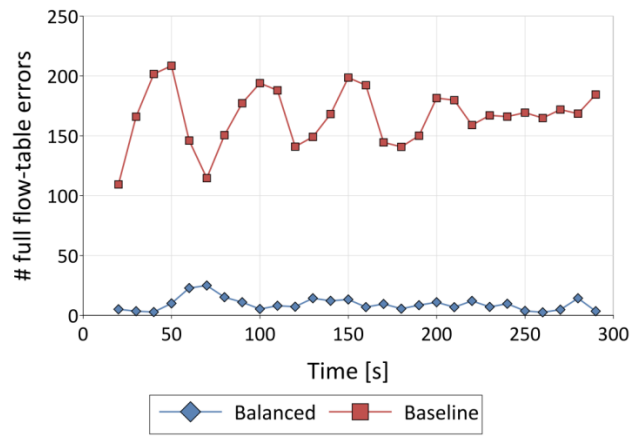

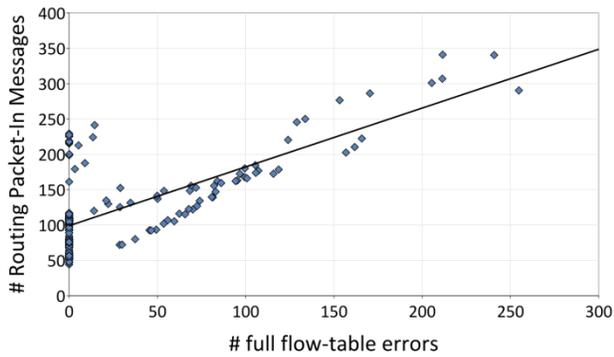

**Figure 12. Control messages as a function of time for ping cycle length of 1 second and flow-table sizes of 1000 entries**

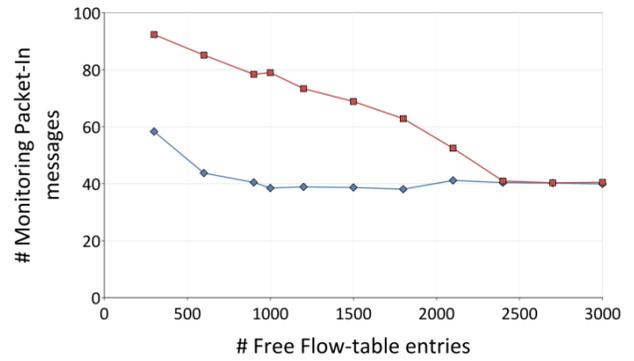

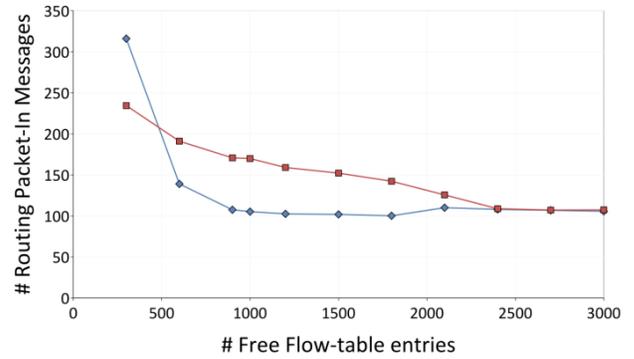

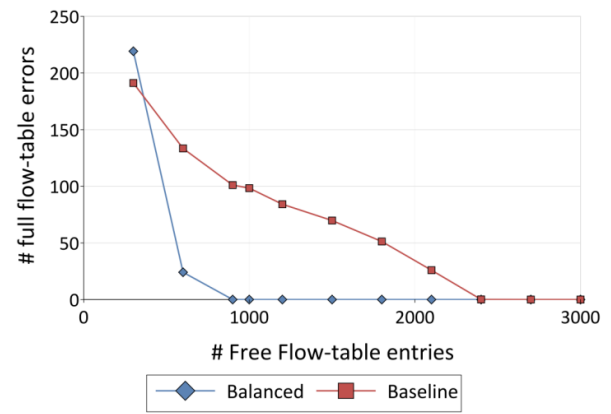

**Figure 14. Control messages vs. the flow-table size for ping cycle of 4 seconds**

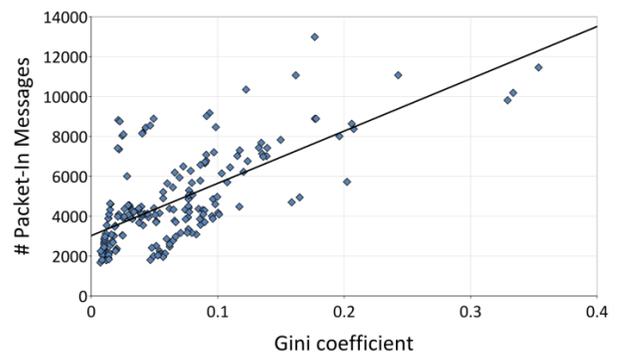

**Figure 13. The number of packet-in messges vs. full flow-table errors**

**Figure 15. Total number of packet in messages vs. the Gini coefficient of free flow-table entries across all switches**

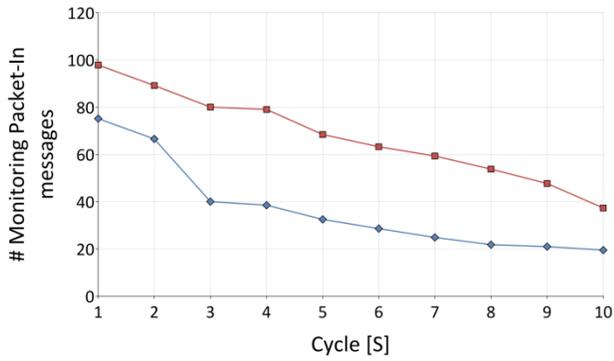

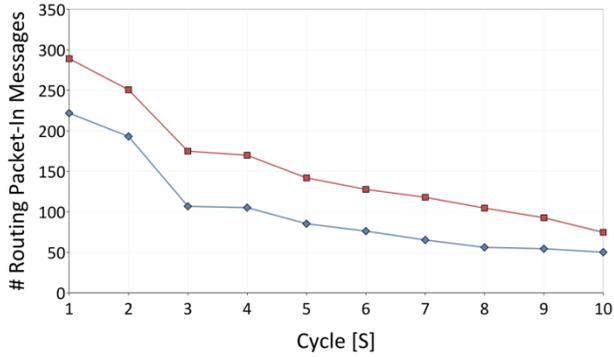

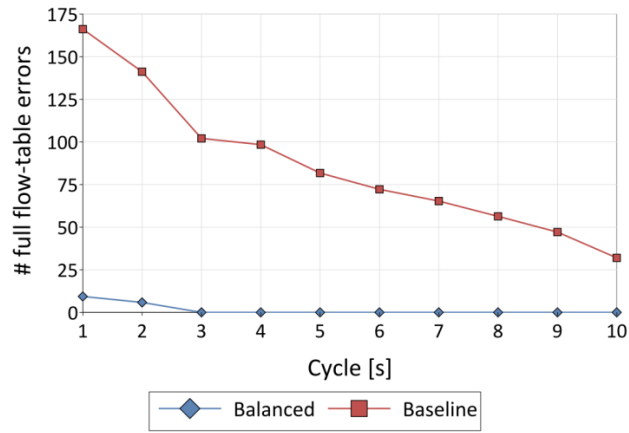

**Figure 16. Control messages vs. ping cycle length for the flow-table size of 1000**

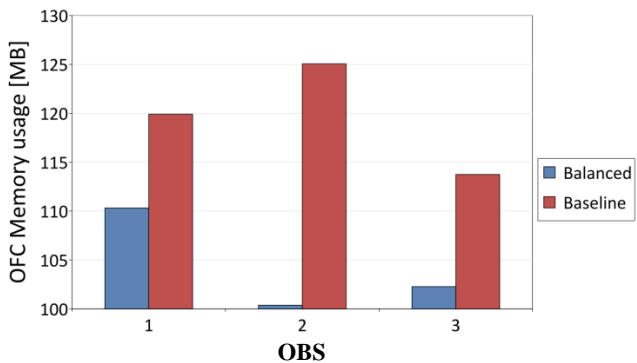

**Figure 17. Average memory usage of the Floodlight controller**

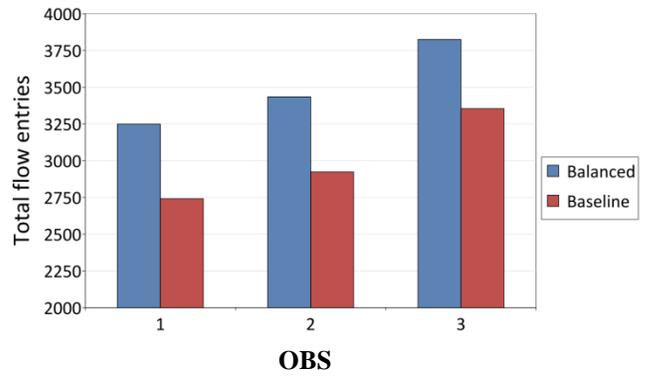

**Figure 18. The total number of used flow entries**

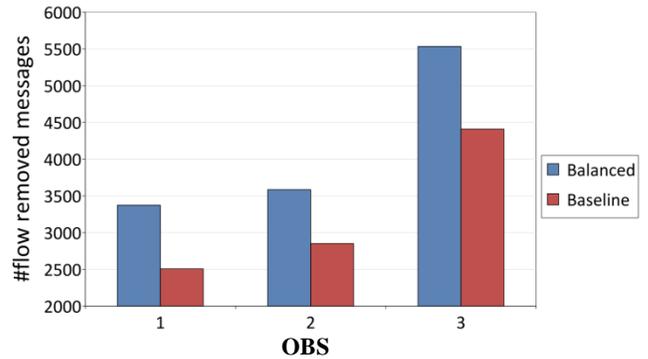

**Figure 19. Amount of collected statistics**

## 6. RELATED WORKS

### 6.1 Monitoring Optimization in OpenFlow

Several recent studies present methods for optimizing statistics collection in OpenFlow. We start with OpenTM [26], which is a method for traffic matrix (TM) estimation in OpenFlow and relies on the pull-based monitoring in contrast to the push-based approach taken in this paper. Nevertheless, the authors conclude that querying the least loaded switch, similar to Algorithm 3, results in the most balanced load. Another major difference between OpenTM and Floware is the optimization objective. OpenTM optimizes the per-flow packet count to account only for packets that reach the destination switches. However, they only consider aggregated flows, while in current research we optimize the monitoring granularity at the IP space.

FlowSense by Yu et al. [21] is a push-based monitoring scheme said to deliver flow statistics at zero cost in contrast to OpenTM. In fact, flow-removed messages, required for collecting the statistics, constitutes a non-negligible overhead. But this overhead is indeed negligible compared to the pull-based approach. In order to keep such a low overhead, FlowSense piggybacks flow-table entries installed by the controller for routing purposes and must not install additional entries. Thus, FlowSence fails to deliver accurate per-flow statistics if aggregated flow-table entries are installed by the controller. The Flow Discovery technique introduced in this paper facilitates per-flow monitoring essential for accurate intrusion detection while avoiding installation of excess flow-table entries.

Recently Chowdhury et al. [22] proposed PayLess, a framework for low-cost statistics collection in OpenFlow. Their adaptive scheduling allows accurate and timely statistics collection, while reducing the network overhead (time dimension). PayLess assumes an external pluggable algorithm that selects flows to be monitored and switches where these flows should be monitored (IP space dimension). Floware's Flow Discovery and Flow Assignment modules can be used as such a plugin in PayLess. Floware assumes a pluggable algorithm which schedules the statistics collection. PayLess' adaptive scheduler can be used as such a plugin in Floware. Thus, Floware and PayLess make a good match, resulting in a framework that optimizes the statistics' collection process in both the IP space and the time dimensions.

## 6.2 Flow Balancing in Cloud Environments

Al-Fares et al. [5] present Hedera, a system which dynamically allocates routes in OpenFlow networks such that the total maximal demand of flows routed through the same link does not exceed the link capacity. Inspired by their work we allocate monitoring resources in Floware such that the total monitoring load does not exceed the capacity of the flow-table.

Another traffic engineering framework, MicroTE, presented by Benson et al. [6] adapts the flow routes to the dynamically changing demands of network flows in order to better utilize the network resources. The authors stress the importance of accurate global view of the traffic matrix. However, they only discuss two alternatives for collecting the flow statistics: polling the switches or allowing the servers to report their traffic demands. The inefficiency of the former alternative was discussed in Section 2.2. The latter alternative requires a special agent to be deployed on each server. Such an agent complicates the data center management and consumes servers' computational resources. MicroTE can be coupled with Floware in order to collect flow statistics from the network fabric with minimal overheads.

## 6.3 Flow Monitoring for Intrusion Detection

Shin and Gu [7] present CloudWatcher monitoring service for dynamic cloud networks where flows are diverted toward a security device for inspection. The authors assume that CloudWhatcher is given the opportunity to decide upon the routes of each individual flow. However, as we discussed in Section 2.2, for the save of efficiency OpenFlow controllers often install aggregated flow-table entries that match multiple flows. In cases where specific short lived flows need to be inspected, our dynamic flow discovery technique can be used to alert CloudWhatcher upon discovery of flows that need to be diverted.

Ballard et al. [33] proposed improving traffic mirroring in OpenFlow and presented OpenSafe, an IDS that analyzes the exported port mirroring statistics according to pre-specified policies. One of the major disadvantages of OpenSafe is the network load overhead that is created by the mirrored traffic.

The InMon Corporation offers hybrid controller with DDoS mitigation [34]. The controller, which makes use of the sFlow and OpenFlow standards, provides real-time detection and mitigation of DDoS attacks. Upon detection the DDoS mitigation SDN application pushes static flow entries to selected switches that drop the attack packets. Switches containing a sFlow agent continuously send traffic measurements to the sFlow-RT controller. sFlow-RT utilizes the flexibility of OpenFlow to mitigate the attacks, but requires special software support for the switches and controller. The primary disadvantage of sFlow-RT and similar solutions is the modifications they introduce into the specification and the implementation of OpenFlow components.

## 7. CONCLUSIONS AND FUTURE WORK

In this paper we introduced Floware, an enabling technology that makes it possible to bring legacy network monitoring solutions to the SDN environment. We demonstrated a complete end-to-end monitoring process starting with optimized selection of observed switches, through Flow Discovery, optimized Flow Assignment, and statistics collection scheduling, and finally ended with the export of NetFlow datagrams to a commercial traffic analysis tool. Although, we refer to NetFlow as a representative protocol for collecting flow statistics in legacy networks the presented framework can be trivially extended to support other protocols as well.

Empirical evaluation suggests that the proposed Flow Assignment algorithm saves precious network resources by balancing the monitoring effort among many network switches. These resources can be used to increase the number and the granularity of monitored flows. Thus traffic analysis tools in general and flow based NIDS in particular, will be able to inspect more traffic and the statistics it receives will be more accurate. The saved resources also reduce the chance for errors and congestions in the network.

OpenFlow is a rapidly changing technology with many challenges that have not yet been addressed. Recently, OpenFlow v1.3 [3] was adopted by major OFCs. OpenFlow v1.3 increases the flexibility of data plane by introducing the flow-table pipeline process. A special, next table, action directs the switch to continue checking for a matching entry in the next flow-table.

Floware can utilize the flow-table pipeline introduced in OpenFlow v1.3 in order to completely decouple flow monitoring and routing decisions. Currently, we must define the next hop in the action field of the installed active flow entries. In the future, the network administrator will be able to dedicate the first flow-table in the pipeline for monitoring purposes.

In such cases, Floware will install every flow-discovery entry with these two actions: send to controller and next table. This will eliminate the delay of traffic flows initiated by using flow-table entries previously installed by the controller (without waiting for the exact match active flow entry to be installed).

## 8. ACKNOWLEDGMENT

The research was sponsored by the Kabarnit Cyber consortium funded by the Chief Scientist in the Israeli Ministry of Economy under the Magnet Program.

We thankfully acknowledge the willingness of the members of the Floodlight mailing list and the NetFlow Analyzer support team to answer our questions.